\documentclass[a4paper,11pt]{article}
\usepackage{helvet}
\usepackage{lmodern}
\usepackage{graphicx}
\usepackage{color}
\usepackage{geometry}
\usepackage{enumitem}
\usepackage{chngcntr}
\usepackage{gensymb}
\usepackage{amsmath}
\numberwithin{equation}{section}
\usepackage{float}
\usepackage{pos}
\usepackage{wrapfig}
\usepackage{subfigure}

\title{Particle density fluctuations and correlations in low energy Cosmic-Ray showers simulated with CORSIKA}
\ShortTitle{Particle density fluctuations and correlations}

\author*[a]{Weronika Stanek}
\author[b]{Jerzy Pryga}
\forColl{CREDO}

\affiliation[a]{AGH University of Science and Technology, Faculty of Physics and Applied Computer Science,\\ Krakow, Poland}

\affiliation[b]{Jagiellonian University, Faculty of Physics, Astronomy and Applied Computer Science,\\ Krakow, Poland}

\emailAdd{wstanek@student.agh.edu.pl}

\abstract{The current studies of cosmic rays are focused on most energetic particles entering the atmosphere and producing a single Extensive Air Shower (EAS). There are, however, models predicting that interactions of high energy particles may result in Cosmic-Ray Ensembles (CRE) created far from the Earth. They could be observed as some number of correlated air showers of relatively low energies spread over a large area. The objective of the Cosmic Ray Extremely Distributed Observatory (CREDO) is to search for CRE using all available data from different detectors and observatories including even small but numerous detectors spread over large areas.\newline

Interpretation of such measurements require precise information on properties of EAS in a very wide energy spectrum. Low energy EAS are analysed using events from CORSIKA, the program performing air shower simulations. The primary cosmic ray particle energy range extends from 1~TeV up to 4\;000~TeV. The secondary particles at the ground level are studied in order to obtain their density fluctuations and correlations in location. Although the fluctuations observed in multiplicity distributions are consistent with random the more detailed analysis reveals that near a selected particle the density of other particles is enhanced over that expected in the absence of correlations. The results of this analysis may be useful in further calculations, for example to obtain probability of detection of an EAS without special simulations.}

\FullConference{37$^{\rm{th}}$ International Cosmic Ray Conference (ICRC 2021)\\
		July 12th -- 23rd, 2021\\
		Online -- Berlin, Germany}

\begin{document}
\maketitle
\section{Introduction} \label{sec0}
Cosmic rays provide access to the highest energies in the Universe. They can be observed indirectly through Extensive Air Showers, which are produced through primary particle interaction with Earth atmosphere \cite{b16}. In most cases air shower is created by a single Cosmic-Ray particle interacting for the first time in the Earth's atmosphere. Although it was not confirmed yet, multiple correlated particles, called Cosmic-Ray Ensembles (CRE), may sometimes reach the Earth.  They may be considered as a group of spatially or temporally correlated particles which decayed from a~common primary or originate from the same interaction somewhere outside the Earth. Secondary particles produced in a Cosmic-Ray Ensemble may cover a large area, even thousands kilometers wide. They initiate local showers which may be detected in several places on Earth and can be associated by their coincidence in time \cite{credo, b21}.

The Cosmic Ray Extremely Distributed Observatory (CREDO) is a global collaboration which aims to observe and analyse Cosmic-Ray Ensembles. Since the CRE cascades may manifest as a bunch of correlated particles spread over large areas, CREDO main objective is to combine data from a large number of detectors. It is willing to include all available data from smartphones, small  scintillator detectors, educational detectors and any professional infrastructure, like Pierre Auger Observatory, Telescope Array, Ice Cube, Baikal-GVD and other observatories and experiments, to study coincidences~\cite{credo, credo_icrc}.

\section{Simulations} \label{sec1}
The knowledge on the low energy cosmic rays as well as the possible features of CRE is still insufficient and it is necessary to examine the theoretical models of Cosmic-Ray cascades \cite{b1}. Searchings for CRE and their reconstruction need a thorough theoretical model of low energy EAS which can be obtained using Monte Carlo simulations \cite{b11}.

CORSIKA (COsmic Ray SImulations for KAscade) is a commonly used program for simulations of EAS initiated by different primary particles with a wide possible range of primary energies. The program is able to recognize 50 particles and nuclei with atomic mass up to 56. All particles contributiong to EAS are tracked in the atmosphere and identified by ID number, trajectory, Lorenz factor, the time since the primary interaction and their Cartesian coordinates \cite{physics, guide}.

In this project 18 sets of cascades were simulated with protons as primary particles. Model used for hadronic interactions at high energies was EPOS-LHC, for hadronic interactions in low energies - URQMD, and for electromagnetic interactions NKG and EGS4 routines. The detailed information can be found in \cite{physics} and \cite{guide}.  

EAS from a single set have their primary particle energy set to a particular value from the range 1~TeV - 4\,000~TeV. The energies are adjusted to include the exact integer powers of 10 and be separated evenly in logarithmic scale. Zenith angle in every case is set to 0. Energy cut for hadrons and muons is 0.3 GeV, for electrons and photons it is 0.003 GeV. However, in this study muons with momentum greater than 1 GeV and electrons and photons with momentum above 0.3 GeV are analysed.

\section{Basic energy dependencies} \label{sec2}
\begin{figure}[H]
	\begin{center}
		\subfigure[]{
			\includegraphics[scale=0.28]{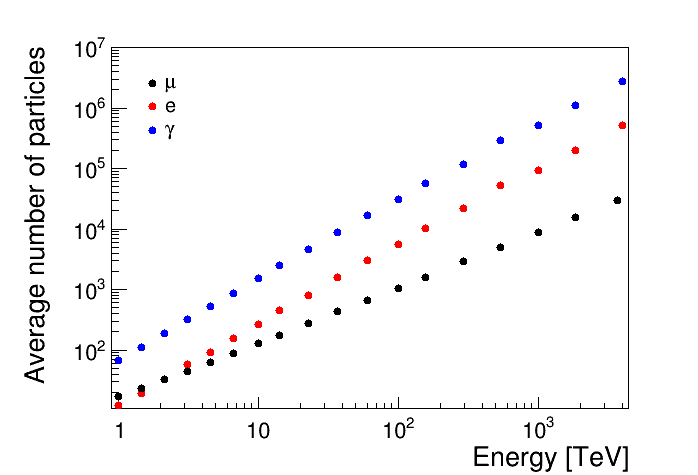}
			\label{R1a}
		}
		\subfigure[]{
			\includegraphics[scale=0.28]{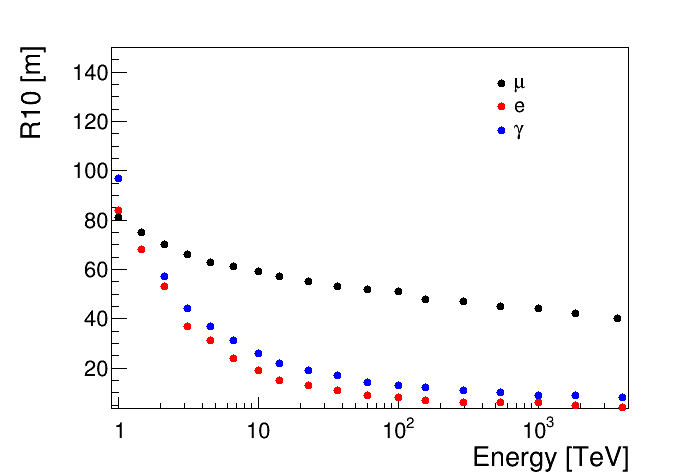}
			\label{R1b}
		}
		\subfigure[]{
			\includegraphics[scale=0.28]{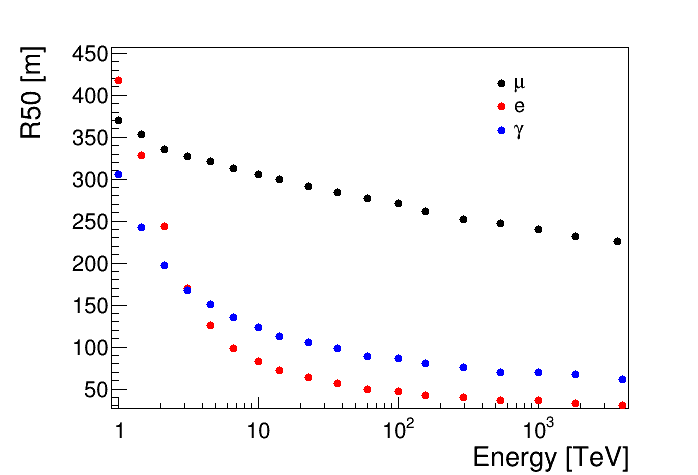}
			\label{R1c}
		}
		\subfigure[]{
			\includegraphics[scale=0.28]{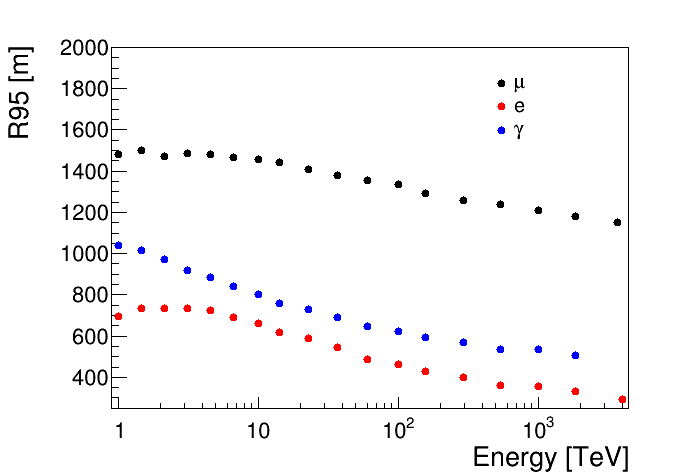}
			\label{R1d}
		}
		\caption{a): Average number of particles as a function of energy of the primary particle; b) - d): Average radius in which 10\% (denoted as $R10$), 50\% ($R50$) and 95\% ($R95$) of particles are included.}
		\label{R1}
	\end{center}
\end{figure}

The first part of the analysis included calculation of the average number of muons, electrons and photons contributing to an EAS with a specific energy as well as calculation of radii of a cascade in which the particular percent of muons, electrons and photons are included. Selected dependencies are shown in Fig.~\ref{R1}.

As it can be expected, the number particle of selected type increases linearly in a $log-log$ scale. Average number of muons increases more rapidly than the number of electromagnetic particles. The $R10$, $R50$, and $R95$ characterize the distribution of particles and also refer to particle density changes in different distances from the centre of an EAS. Interestingly, in every case these radii decrease with the primary energy, although simultaneously the number of particles becomes larger. It clearly indicates that the particles are not distributed evenly but are rather grouped near the axis of a cascade. It occurs mainly due to the fact that with increasing energy, more particles with very small angles are produced and thus the centre becomes denser.

\section{Two particles correlations in location}
\begin{figure}[H]
    \centering
    \includegraphics[scale=0.6]{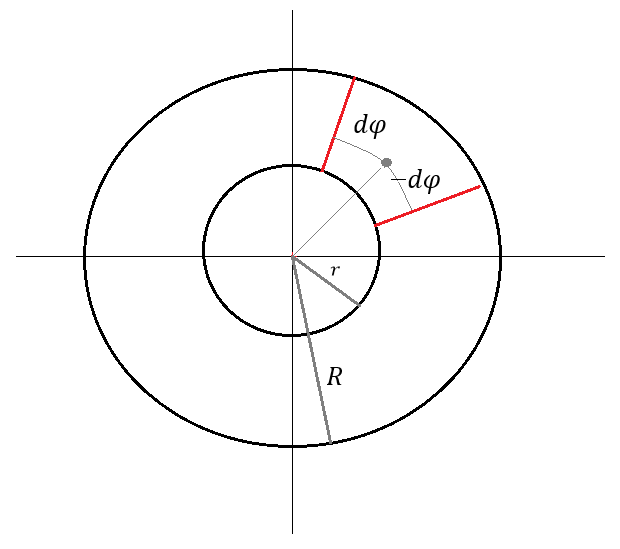}
    \caption{Simple illustration of division of the area covered by a cascade  into rings. The further rings with larger distances from the centre are defined analogically.}
    \label{division}
\end{figure}

Particles contributing to an EAS are not distributed evenly - their density decreases further from the centre as it was shown in previous section. The main objective of the second part of the analysis is to investigate if this is the only dependence on location or if muons, electrons and photons form clusters in which the density is larger than the average. In this analysis photons and electrons are calculated together and denoted as EM particles. The general procedure is as following:

\begin{itemize}
    \item The surface covered by a cascade is divided into rings with a particular width (2~m, 5~m or 20~m). For simplicity, the circle in the middle is not taken into account. 
    
    \item For each particle with coordinates ($r_0$, $\varphi_0$) the neighborhood is defined as a part of the ring (with inner and outer radius,  $r \le r_{0} < R$, respectively) with azimuthal angles in the range ($\varphi_0 - d\varphi$; $\varphi_0 + d\varphi$), $d\varphi = (R-r)/2r$, with the area approximately $\Delta R \times \Delta R$, as shown  in Fig.~\ref{division}. 
    
    \item Particle density is calculated in each ring as well as in each neighborhood area, separately for EM particles (electrons and photons) and for muons. 
\end{itemize}

Ilustration of division of the cascade area is presented in Fig.~\ref{division}. The $d\varphi$ angle is denoted as:
\begin{equation} \label{eq1}
d\varphi = \frac{R-r}{2r}
\end{equation}
where $R$ is the outer radius and $r$ is the inner radius of the ring. In this case, every sector containing a particle is similar to the square $(R-r)\;\times\;(R-r)$.

\begin{figure}[H]
	\begin{center}
		\subfigure[$\Delta R$ = 2~m]{
			\includegraphics[scale=0.28]{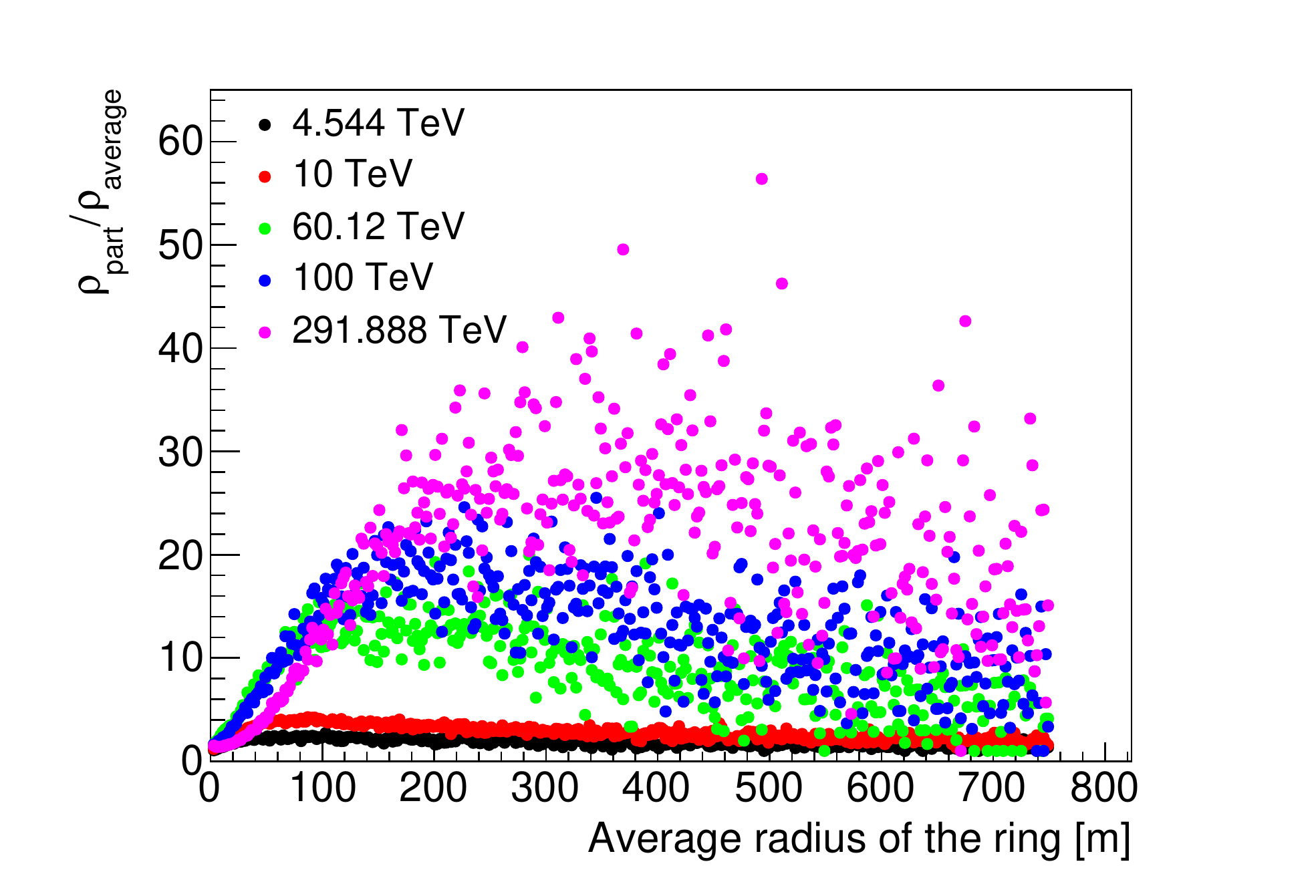}
			\label{R2a}
		}
		\subfigure[$\Delta R$ = 2~m]{
			\includegraphics[scale=0.28]{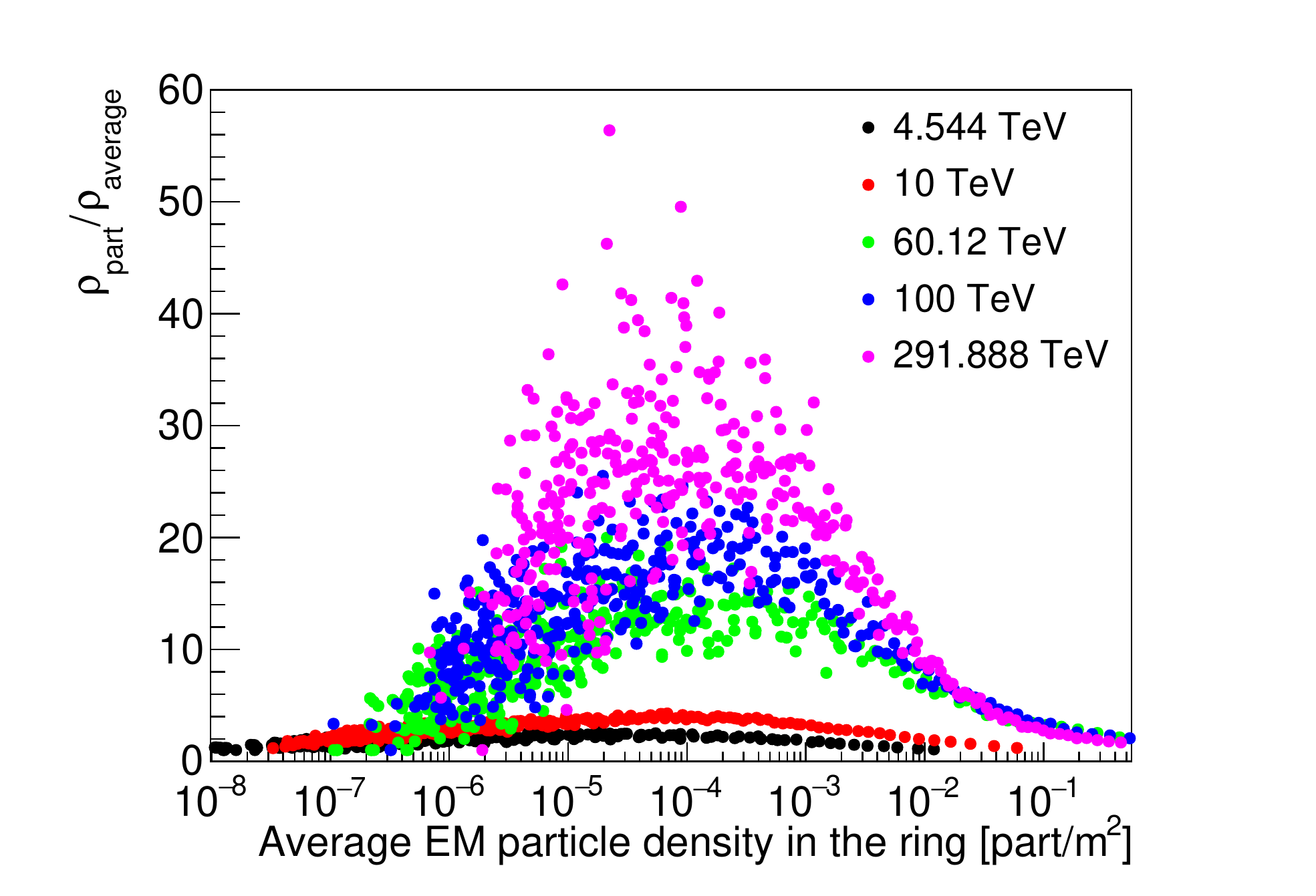}
			\label{R2b}
		}
		\subfigure[$\Delta R$ = 5~m]{
			\includegraphics[scale=0.28]{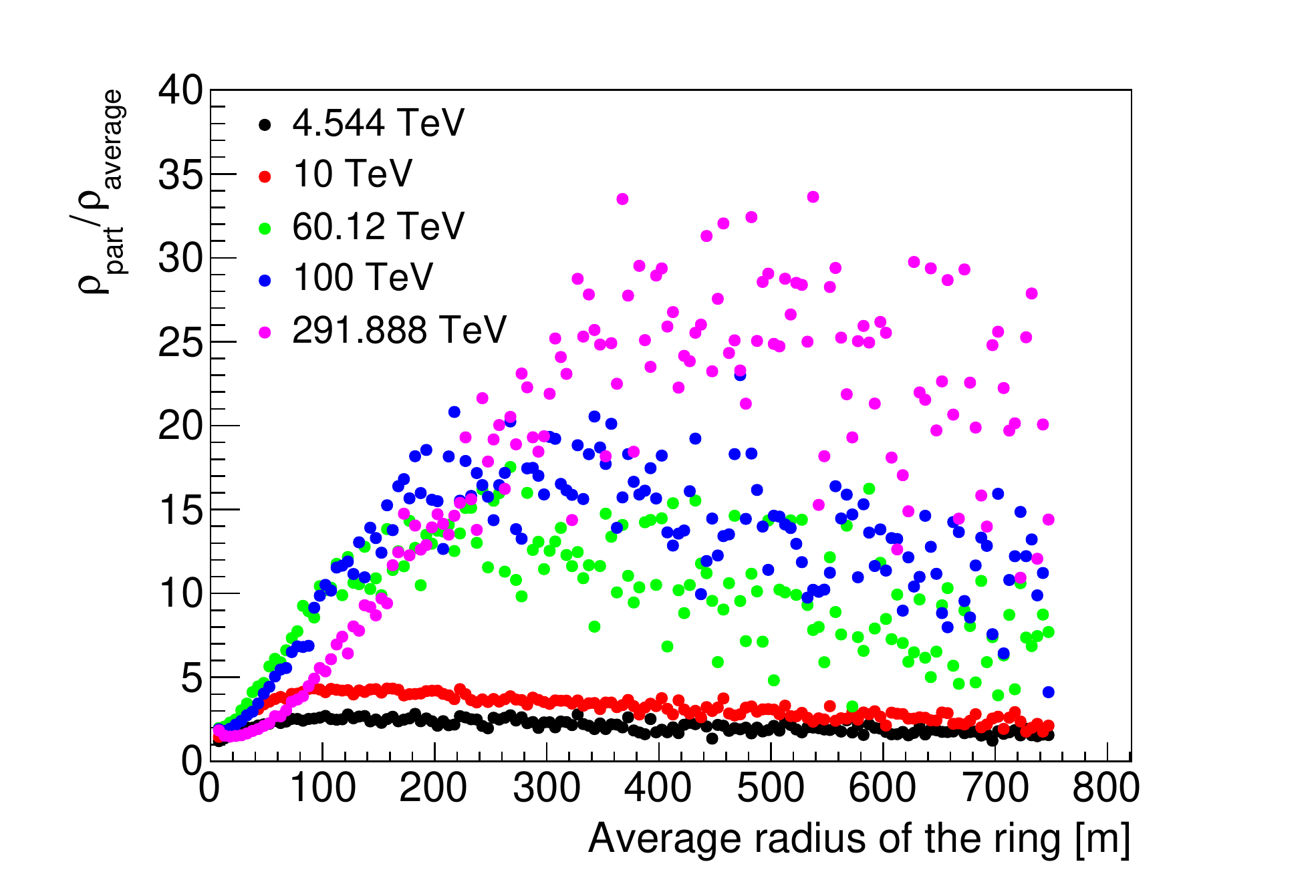}
			\label{R2c}
		}
		\subfigure[$\Delta R$ = 5~m]{
			\includegraphics[scale=0.28]{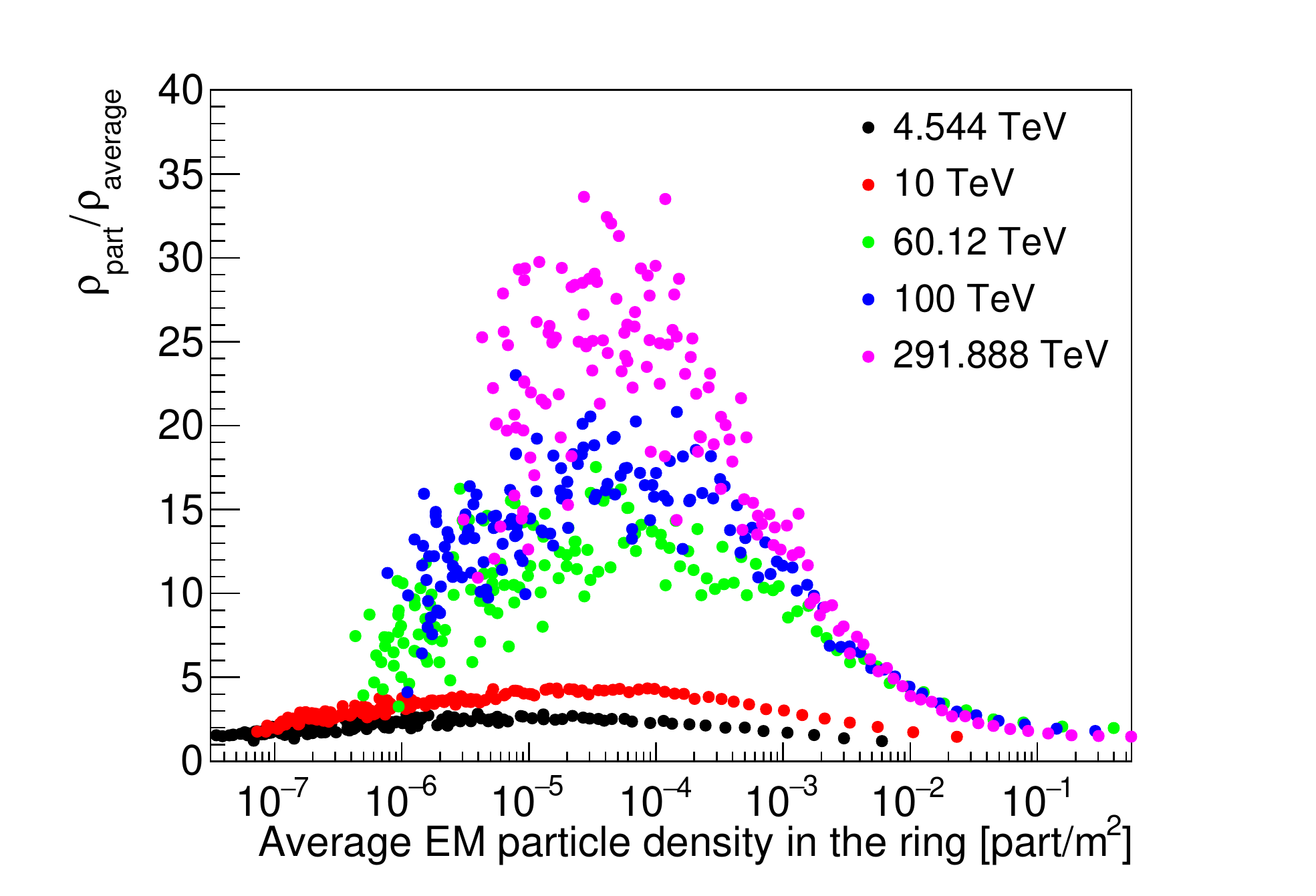}
			\label{R2d}
		}
		\subfigure[$\Delta R$ = 20~m]{
			\includegraphics[scale=0.28]{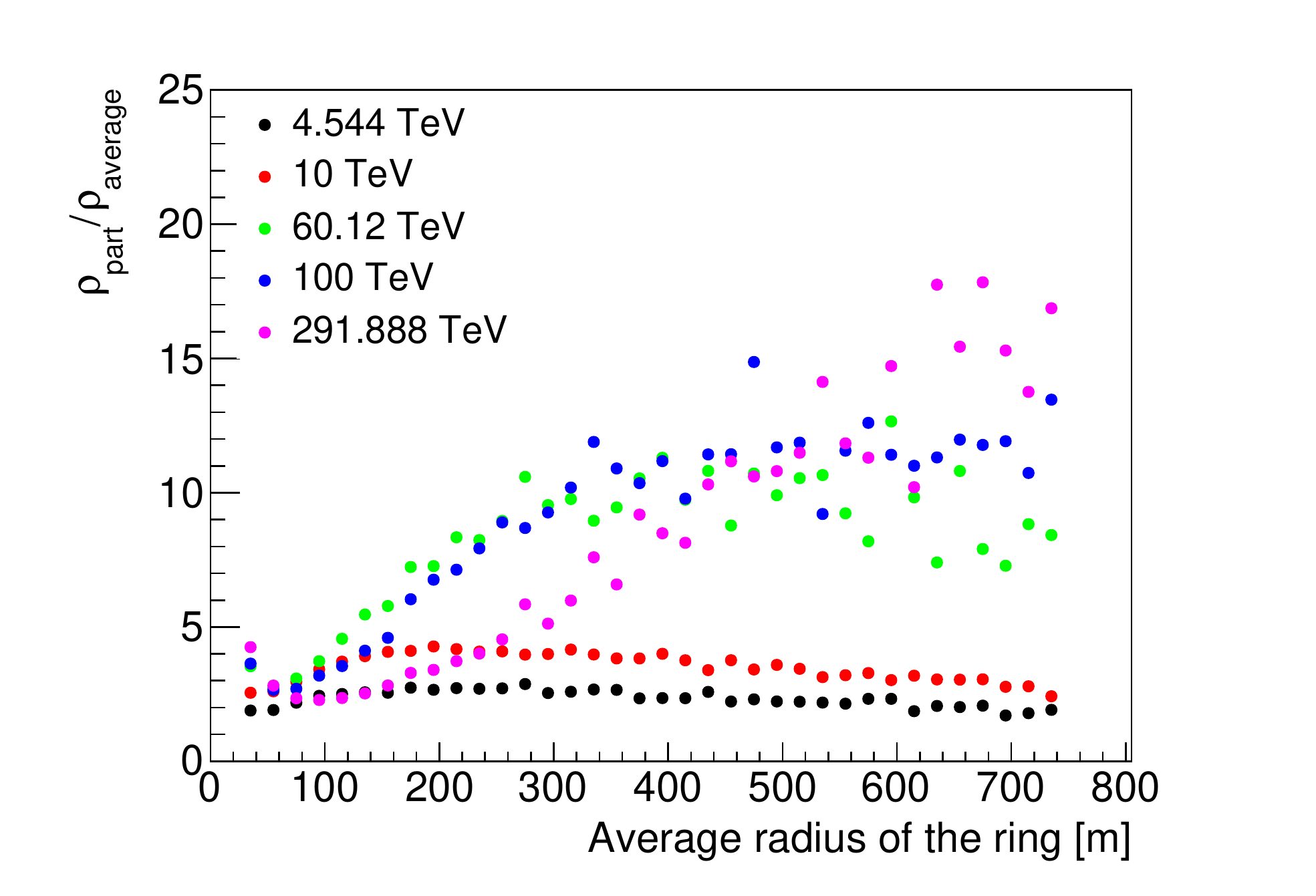}
			\label{R2e}
		}
		\subfigure[$\Delta R$ = 20~m]{
			\includegraphics[scale=0.28]{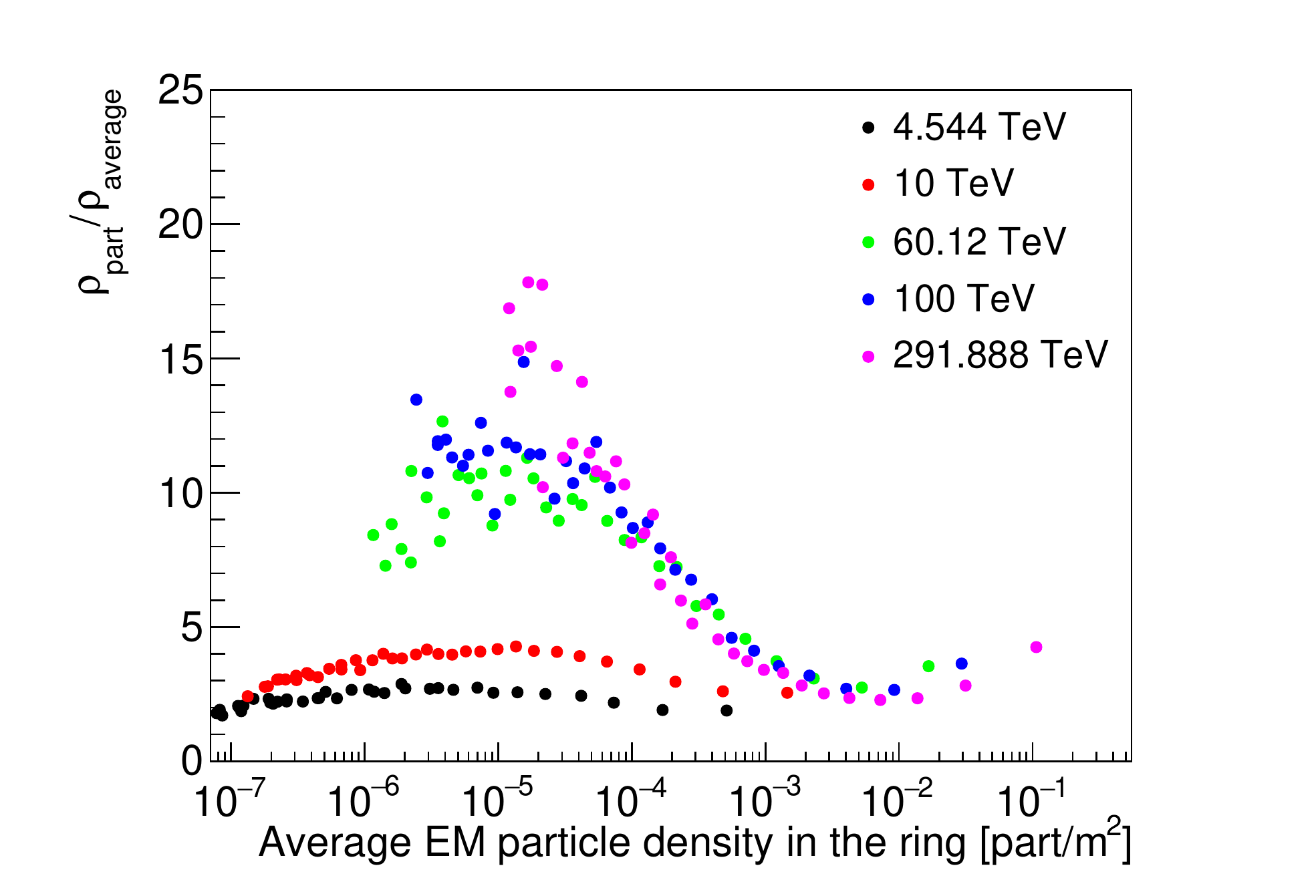}
			\label{R2f}
		}
		\caption{Ratio of density of EM particles in the neighborhood of a selected particle ($e$ or $\gamma$) and the average density in the ring for rings with different values of $\Delta R$.}
		\label{R2}
	\end{center}
\end{figure}

\textbf{EM particles:} Calculations for the EM particles were performed for five primary proton energies: 4.544~TeV, 10~TeV, 60.12~TeV, 100~TeV and 291.888~TeV. The maximum radius of a~cascade was adjusted to 750~m, in order to include 95\% of particles (according to Fig.~\ref{R1d}). Three chosen neighborhoods were 2m $\times$ 2m, 5m $\times$ 5m, 20m $\times$ 20m. Results are presented in Fig.~\ref{R2} as a EM particle density ratio (average density of EM particles in the neighborhood of a selected EM particle, divided by average EM particle density in the whole ring). 

Regarding to Fig.~\ref{R2a}, \ref{R2c}, \ref{R2e}, density ratio is the smallest close to the centre of a cascade and it increases further away from it. Altough the average density decreases, the clustering effect is easily noticeable in larger distances from the centre of the cascade. In the interactions in the atmosphere,  each electron or photon with high or even moderate energy is able to produce a ''subshower'' with an axis different than that of the cascade. The density in the centre of such ''subshower'' is larger than the average. Closer to the centre of the cascade many particles may create their ''subshowers'' and particles from different clusters overlap. In such case, the contributions from correlations present within a single cluster may be smaller than purely random density fluctuations.

The pictures are consistent with results shown in Fig.~\ref{R2} in the right column, where the highest values of density ratio are reached in areas with small average density in a ring.

\begin{figure}[H]
	\begin{center}
		\subfigure[]{
			\includegraphics[scale=0.28]{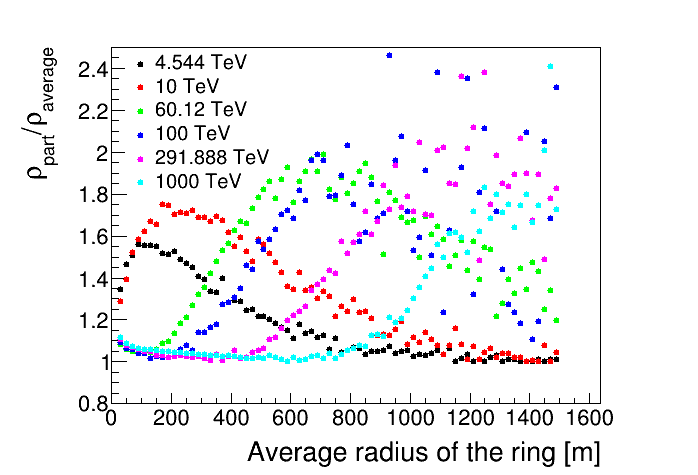}
			\label{R3a}
		}
		\subfigure[]{
			\includegraphics[scale=0.28]{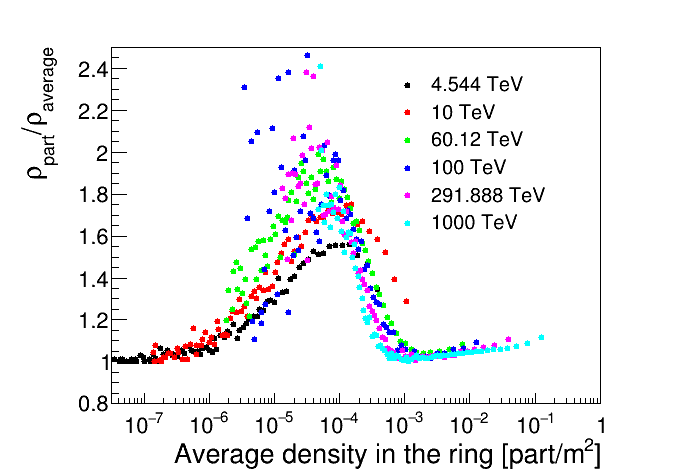}
			\label{R3b}
		}
		\caption{Ratio of density of muons in the neighborhood of a selected muon and the average density in the ring for rings with $\Delta R$ = 20~m.}
		\label{R3}
	\end{center}
\end{figure}

\textbf{Muons:} The analogous analysis only for muons is presented below. The calculations in this section were carried out for six primary particle energies: 4.544~TeV, 10~TeV, 60.12~TeV, 100~TeV, 291.888~TeV and 1000~TeV. The maximum radius was adjusted to 1500~m in order to include 95\% of muons. Results are presented in Fig.~\ref{R3a} and Fig.~\ref{R3b}.

For lower energies (4.544~TeV, 10~TeV) the density ratio achieves the maximum value close to the centre (at $\sim$100~m and $\sim$200~m, respectively). At such place the probability of finding an additional muon near the selected one is approx. 1.6--1.8 times larger than probability of finding a muon in randomly selected place at the same radius. For higher energies the ratio is close to one in the center, but increases farther away from it. For the energy 100 TeV the density ratio is dominated by fluctuations. 

In Fig.~\ref{R3b}, dependences are similar for every energy. There is a peak at the density close to  $10^{-4}$ m$^{-2}$. For densities lower than  $10^{-6}$ m$^{-2}$ there are no  clusters or they are much larger than the area used in the analysis. For rings with densities larger than $10^{-4}$ m$^{-2}$, no clustering effect is visible as muons most probably originate from many clusters. Therefore, the correlations within the cluster become not essential. Only when the density is about $10^{-4}$ m$^{-2}$, the number of muons belonging to a single cluster of correlated particles becomes significantly larger than the number of other muons.

\section{Conclusions}
\begin{itemize}
    \item In the $log-log$ scale the number of EM paricles and muons increases linearly with energy. Radii in which 10\%, 50\%, 95\% of EM particles and muons are included decrease with energy because of kinematics of secondary interactions. At higher energies the emission angles of secondaries are smaller and particles are more and more grouped in the centre.
    \item Results obtained during correlation analysis for EM particles and muons confirm that the density increases around a selected particle and the clustering effect is visible.
    \item Probability of finding a cluster of EM particles is the largest at some distance from the cascade center and the particle density may be enhanced there by a factor $\sim$ 20 or even more.
    \item In case of muons, the density ratio reaches the maximum value at different radius depending on energy but in these places the average density is similar for all energies. In such areas the probability of finding an additional muon near the selected one is about 2 times larger than in other areas of the ring.
\end{itemize}

\section*{Acknowledgements}
This study has been supported by PLGrid Infrastructure. We would like to thank the ACC Cyfronet AGH-UST for their supercomputing support.

\clearpage
\section*{Full Authors List: \Coll\ Collaboration}
%
%
\scriptsize
\noindent
Weronika Stanek$^1$,
Jerzy Pryga$^2$,
Krzysztof W. Wo{\'z}niak$^3$,
Piotr Homola$^3$, 
David E. Alvarez Castillo$^{3,4}$, 
Dmitriy Beznosko$^5$,
Nikolai Budnev$^6$,
Dariusz G{\'o}ra$^3$,
Alok C. Gupta$^7$,
Bohdan Hnatyk$^8$,
Marcin Kasztelan$^9$,
Peter Kovacs$^{10}$,
Bartosz {\L}ozowski$^{11}$,
Mikhail~V.~Medvedev$^{12,13}$, 
Justyna Miszczyk$^3$,
Alona Mozgova$^8$,
Vahab Nazari$^{4,3}$, 
Micha\l{} Nied{\'z}wiecki$^{14}$,
Mat{\' i}as Rosas$^{15}$,
Krzysztof Rzecki$^{1}$,
Katarzyna Smelcerz$^{14}$,
Karel Smolek$^{16}$,
Jaros\l{}aw Stasielak$^{3}$,
S\l{}awomir Stuglik$^{3}$,
Oleksandr Sushchov$^3$, 
Arman Tursunov$^{17}$,
Tadeusz Wibig$^{18}$, 
Jilberto Zamora-Saa$^{19}$. \\ 

\noindent
$^{1}$AGH University of Science and Technology, Mickiewicz Ave., 30-059 Krak{\'o}w, Poland.\\
$^2$Jagiellonian University, Go{\l}ębia 24, 31-007 Kraków. \\
$^3$Institute of Nuclear Physics Polish Academy of Sciences, Radzikowskiego 152, 31-342 Krak{\'o}w, Poland.\\
$^4$Joint Institute for Nuclear Research, Dubna, 141980 Russia.\\ 
$^5$Clayton State University, Morrow, Georgia, USA.\\
$^6$Irkutsk State University, Russia.\\
$^7$Aryabhatta Research Institue of Observational Sciences (ARIES), Manora Peak, Nainital 263001, India.\\
$^8$Astronomical Observatory of Taras Shevchenko National University of Kyiv, 04053 Kyiv, Ukraine.\\
$^9$National Centre for Nuclear Research, Andrzeja Soltana 7, 05-400 Otwock-{\'S}wierk, Poland.\\
$^{10}$Institute for Particle and Nuclear Physics, Wigner Research Centre for Physics, 1121 Budapest, Konkoly-Thege Mikl{\'o}s {\'u}t 29-33, Hungary.\\
$^{11}$Faculty of Natural Sciences, University of Silesia in Katowice, Bankowa 9, 40-007 Katowice, Poland.\\
$^{12}$Department of Physics and Astronomy, University of Kansas, Lawrence, KS 66045, USA.\\
$^{13}$Laboratory for Nuclear Science, Massachusetts Institute of Technology, Cambridge, MA 02139, USA.\\
$^{14}$Department of Computer Science, Faculty of Computer Science and Telecommunications, Cracow University of Technology, Warszawska 24, 31-155  Krak{\'o}w, Poland.\\
$^{15}$Liceo 6 Francisco Bauz{\' a}, Montevideo, Uruguay.\\
$^{16}$Institute of Experimental and Applied Physics, Czech Technical University in Prague.\\
$^{17}$Research Centre for Theoretical Physics and Astrophysics, Institute of Physics, Silesian University in Opava, Bezru{\v c}ovo n{\'a}m. 13, CZ-74601 Opava, Czech Republic.\\
$^{18}$University of {\L}{\'o}d{\'z}, Faculty of Physics and Applied Informatics, 90-236 {\L}{\'o}d{\'z}, Pomorska 149/153, Poland.\\
$^{19}$ Universidad Andres Bello, Departamento de Ciencias Fisicas, Facultad de Ciencias Exactas, Avenida Republica 498, Santiago, Chile.\\

\end{document}